# Microwave-acoustic-based isolated gate driver for power electronics

Liyang Jin[1], Zichen Xi[1], Joseph G. Thomas[1,2], Jun Ji[1], Yuanzhi Zhang[1], Nuo Chen[1], Yizheng Zhu[1,2], Linbo Shao[1,3,*] and Liyan Zhu[1,4,*]

[1]Bradley Department of Electrical and Computer Engineering, Virginia Tech, Blacksburg, VA 24061, USA.
[2]Center for Photonics Technology, Virginia Tech, Blacksburg, VA 24061, USA.
[3]Department of Physics and Center for Quantum Information Science and Engineering (VTQ), Virginia Tech, Blacksburg, VA 24061, USA.
[4]Center for Power Electronics Systems (CPES), Virginia Tech, Blacksburg, VA 24061, USA.
* Corresponding authors: shaolb@vt.edu; liyanz@vt.edu

## Abstract

Electrical isolation is critical to ensure safety and minimize electromagnetic interference (EMI), yet existing methods struggle to simultaneously transmit power and signals through a unified channel. Here we demonstrate a mechanically-isolated gate driver based on microwave-frequency surface acoustic wave (SAW) device on lithium niobate that achieves galvanic isolation of 2.75 kV with ultralow isolation capacitance (0.032 pF) over 1.25 mm mechanical propagation length, delivering 13.4 V open-circuit voltage and 44.4 mA short-circuit current. We demonstrate isolated gate driving for a gallium nitride (GaN) high-electron-mobility transistor, achieving a turn-on time of 108.8 ns comparable to commercial drivers and validate its operation in a buck converter. In addition, our SAW device operates over an ultrawide temperature range from 0.5 K (-272.6 °C) to 544 K (271 °C). The microwave-frequency SAW devices offer inherent EMI immunity and potential for heterogeneous integration on multiple semiconductor platforms, enabling compact, high-performance isolated power and signal transmission in advanced power electronics.

## Main

Power electronics form the backbone for modern electrical and electronic systems, supporting the applications such as renewable energies, data centers, electric vehicles and space explorations[1,2]. These systems usually operate across different voltage levels, including high voltages for power stages, and low voltages for digital computing, control, and sensing stages. For example, the large DC motors used on high-speed trains can take 6.6 kV and 150 A[3,4], while their control and sensing circuits remain at low voltages (typically around 1 to 5 V)[5,6]; data centers receives medium-voltage AC (up to 35 kV) from utilities and step it down through multiple stages to GPUs (a typical core voltage is around 1 V)[7,8] . Such large voltage differences across high-power circuits and sensitive circuits necessitate robust and efficient electrical isolation for preventing fault propagation, ensuring user and equipment safety, and preserving signal integrity in complex electromagnetic environments[9-12]. This isolation typically requires separating both control signals and the power that drives them, making simultaneous co-transmission of isolated power and high-speed signals an attractive capability. As a representative example of isolated power and signal co-transmission, gate drivers for high-voltage transistors, which act as critical interfaces between control and power domains, typically require both isolated pulse-width modulation (PWM) signals and isolated power supplies to deliver 5–20 V gate control signals. This is usually implemented using discrete digital isolators and isolated power modules, which increases system complexity and vulnerability[13,14].

Galvanic isolated gate drivers have been achieved by inductive[15-17], capacitive[18-20], optoelectronic components[21-23] (**Figs. 1a and 1b**) and low-frequency piezoelectric transducers[24,25] (**Fig. 1c**). Inductive transformers offer high power capability and efficient energy conversion. However, they are typically bulky, emit electromagnetic radiation, and require separate channels for signal and power transmission[26]. In addition, their large interwinding capacitance can act as a path for electromagnetic interference (EMI) propagation. Capacitive isolation offers high-bandwidth signal transmission, but the fragile on-chip capacitive barriers are typically susceptible to EMI. Besides, there is currently no simple and cost-effective solution for high-power co-transmission using capacitive methods. Benefiting from the hundreds of nanometers optical wavelength, which is orders of magnitude shorter than that of kHz to MHz electrical signals used in power systems, optical isolators (also known as, optocoupler or photocoupler) are naturally immune to EMI but need conversions between electrical and optical domains. Photovoltaic converters can transform light into electricity, therefore deliver energy, but their slow response - due to the large device area and high junction capacitor in photovoltaic region - makes them not suitable for transferring high-frequency signals[21]. Photodiodes under reverse bias can detect light modulated signals at frequencies over 10s GHz, but their typical operation photocurrent is at or below mA regions[27]. Moreover, the reverse bias voltage needs to be provided by a separate isolated power supply in power applications[23]. In summary, current inductive, capacitive and optical technologies for effective signal transmissions still require parallel isolated power supplies. This typically involves inefficiently isolated DC-DC converter, which increases component count, occupies precious board space, compromises power density, and acts as an additional source of EMI[28].

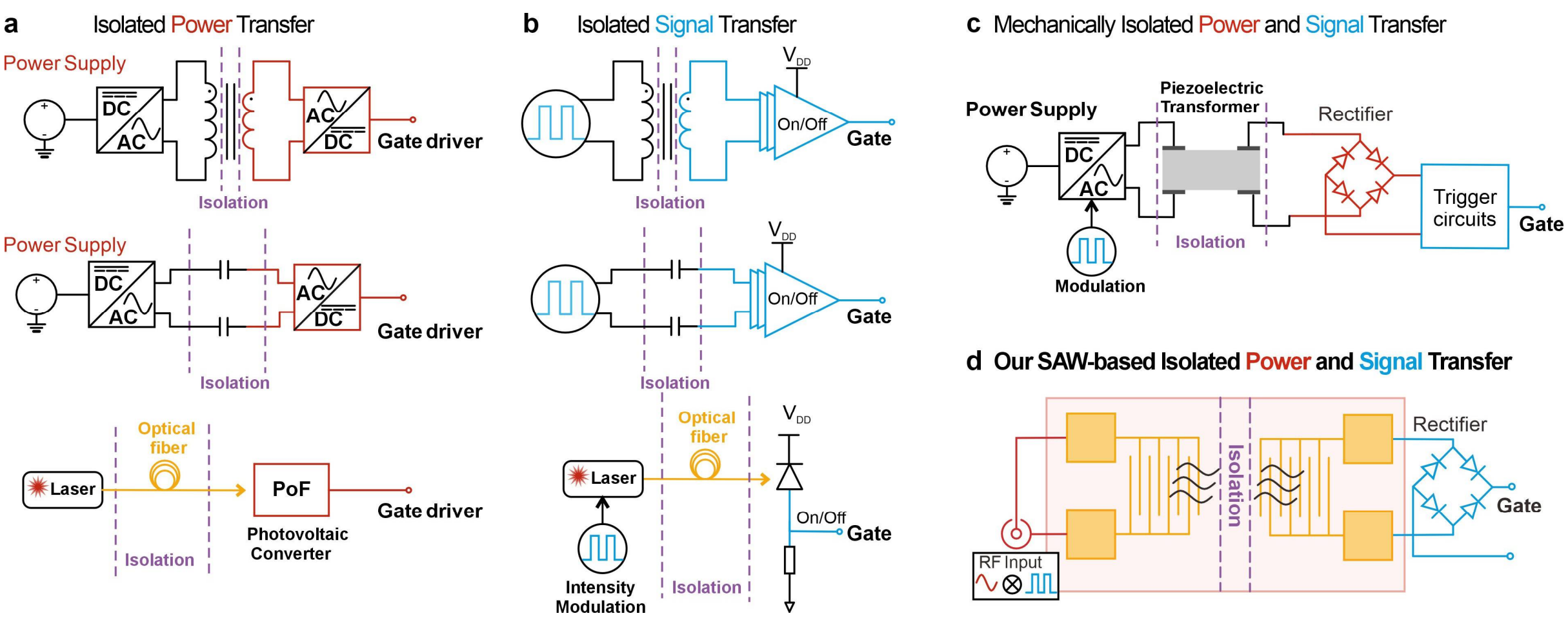

**Fig. 1 | Overview of galvanic isolation principles in power electronic control applications. a,** Isolated power and **b,** Signal transfer using inductive, capacitive and optical coupling methods. **c,** Low-frequency piezoelectric transformer for unified power and signal transfer. **d,** Our microwave-frequency SAW-device based power and signal co-transmission in gate driver application.

Meanwhile, bulk-mode piezoelectric transformers are used to unify power and signal transfer (**Fig. 1c**)[24,25], featuring excellent EMI immunity by their mechanical nature. Current piezoelectric transformers use low-frequency bulk mechanical (elastic) waves generated piezoelectrically to deliver power from the primary to the secondary side. At the receiver end, passive components, such as diode bridge rectifiers and trigger circuits, are used for signal reconstruction, providing high fidelity signals with power driving capabilities[24]. Nevertheless, for current piezoelectric transformers, their operating frequencies below tens of megahertz and their high mechanical $Q$ factors of about 1,000 results in an inherent narrow bandwidth of tens kilohertz, insufficient for advanced gate driving in wide-bandgap (WBG) power electronics, where sub-micro-second transitions are required.

In this Article, we demonstrate isolated power and signal co-transmission using a small-footprint microwave-frequency surface acoustic wave (SAW) device on a lithium niobate (LN, $LiNbO_3$) substrate (**Fig. 1d**). The device with an on-chip mechanical propagation length of 1.25 mm achieves a galvanic isolation of 2.75 kV and an ultralow isolation capacitance of 0.032 pF. Using interdigital transducers (IDTs) to bidirectionally convert energy and signals between electrical and mechanical domains, our SAW device provides an open-circuit voltage of 13.4 V and a short circuit current of 44.4 mA at the receiver end. We prototype an isolated gate driver for a 650-V, 11-A gallium nitride (GaN) power high-electron-mobility transistor (HEMT), achieving a turn-on time of 108.8 ns, which is a significant improvement compared to previous exploration on gate drivers based on SAW devices[29,30]. We recognize the gap in switching times between our SAW gate driver and commercial gate driver ICs, we highlight that this is a mechanical isolation approach, proving ultralow isolation capacitance and EMI immunity. We also demonstrate a buck converter to showcase the device's capability of driving high-side floating gates. Furthermore, we demonstrate an ultrawide operational temperature range of our SAW devices, from 0.5 K (-272.6 °C) to 544 K (271 °C).

## Results

### Device characterization and application in gate driving

Our microwave-frequency SAW device (**Fig. 2a**) consists of two pairs of periodically spaced, cross-finger electrodes, known as IDTs, to piezoelectrically generate and receive microwave-frequency acoustic waves. The detailed device design is discussed in Methods and **Supplementary Fig. 1**. A full-bridge rectifier is used to down-convert the received signal from microwave frequency to low frequency; for example, from amplitude-modulated 223 MHz signals to 10s kHz square wave signals. The maximum cable-to-cable power transmission is -5.12 dB for the 223 MHz acoustic wave, and -8.07 dB if time gated to the single-pass acoustic propagation (**Fig. 2b** and **Supplementary Fig. 2**). We attribute 6 dB of this insertion loss to the IDT symmetric design and the remaining to the resistive loss and slight impedance mismatch of IDTs. The variance in the transmission spectrum is due to the reflections between IDTs, which does not affect the gate driving performance. We note that the maximum $S_{21}$ transmissions show small differences between the SAW devices with different acoustic-wave propagation distances. The propagation loss is expected be less than 0.2 dB/mm for a 223-MHz SAW on 128°Y-cut LN (assuming constant frequency qualify factor *fQ* product).

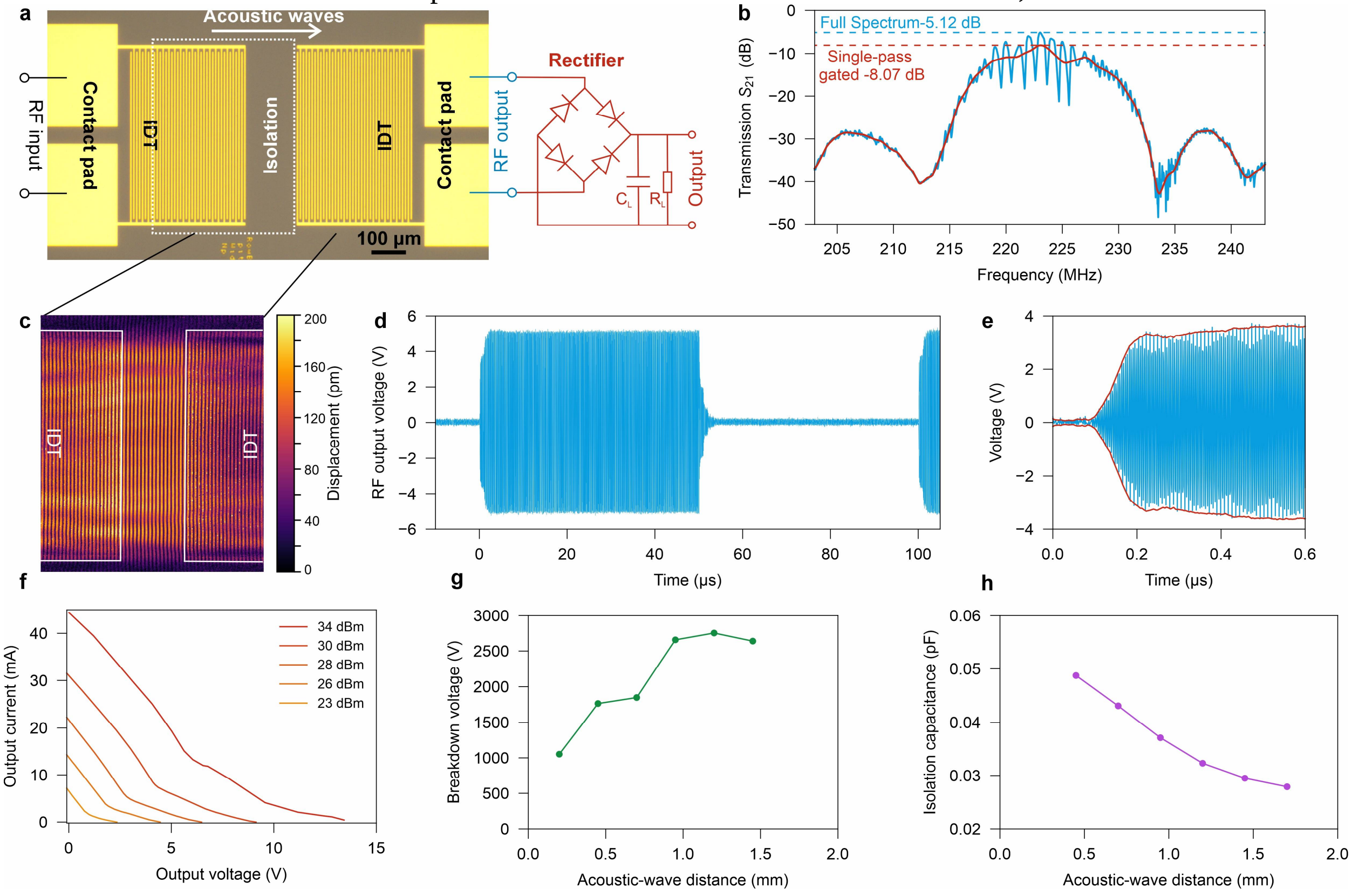

**Fig. 2 | Design and characterization of SAW gate driver. a,** Microscope image of IDTs and simplified circuit schematic of gate driver setup. **b,** $S_{21}$ spectrum of the SAW device featuring an acoustic mode at 223 MHz with -5.12 dB maximum power transmission. The red curve shows the $S_{21}$ spectrum with time gating to the single-pass signal. **c,** Vibration amplitude scanning of the IDT device. The input RF signal is at frequency of 223 MHz and with a power of 23 dBm. **d,** 10 kHz modulated RF output signal at receiver end. **e,** Zoomed-in rising edge of the output waveform. **f,** I-V characteristics of the SAW device under different input power **g,** SAW device breakdown voltages at different IDT gaps. **h,** SAW device isolation capacitances at different IDT gaps.

Leveraging our spectrometer-based optical vibrometer[34], we measure the mechanical displacement profile showing the acoustic wave propagation between the input and output IDTs (**Fig. 2c**). High-frequency SAWs are well capable of carrying energies in solids; the SAWs are well confined in the device region with minimal leakages. This allows future designs to put multiple SAW devices on a single chip die without crosstalk.

Our SAW device achieves electrically isolated power and signal co-transmission for gate driving of power transistors, while only signal transmission through SAW devices was explored in previous work[35]. The carrier frequency of the radio frequency (RF) signal is tuned to the optimal frequency of the IDT. A 10 kHz, 50% duty cycle pulse-width modulation (PWM) waveform (**Fig. 2d**) is used to modulate the RF carrier signals, leading to the RF output with a rise time of 75.05 ns (**Fig. 2e**), which is sufficiently fast for gate driving of power transistors. The full-bridge rectifier converts the received RF output back into an electrical PWM signal for given load (i.e. the gate of a GaN HEMT) which enables simultaneous electric power and encoded control signal transmission. The impacts of signal rise time by the SAW device and the rectifier are further investigated in Supplementary Fig. 3.

We characterize output performance of our device by measuring its current-voltage (I-V) response under various load conditions (**Fig. 2f**). Our SAW device behaves like a voltage source with a slightly nonlinear internal resistance, providing an open-circuit voltage of 13.4 V and a short circuit current of 44.4 mA at the receiver end, which eliminates the need for separate bulky isolated power supplies.

We characterize the isolation performance of our devices. The breakdown voltages of our SAW devices increase with increasing acoustic-wave propagation distances, *i.e.*, larger gaps between input and output IDTs (**Fig. 2g**). An acoustic-wave distance of 1 mm or greater allows our SAW gate driver to achieve a breakdown voltage exceeding 2.7 kV. We also characterize the isolation capacitances of devices with different acoustic-wave propagation distances (**Fig. 2h**), which could be the potential propagation path for common-mode current induced EMI noises. We measured an isolation capacitance of 0.032 pF at distance of 1.25 mm, reduced by an order of magnitude compared to previous isolation solutions[36]. This ultralow isolation capacitance results from the relatively large physical distance between IDTs and the small footprint (100s μm by 100s μm) of an individual IDT. Overall, our SAW gate driver has the potential to provide inherent multi-kilovolt galvanic isolation with intrinsic immunity to EMI.

### Microwave acoustic gate driving of GaN HEMT

We evaluate the gate driving performance of our SAW device by the double pulse test (DPT), which is widely used to characterize the transient dynamics of power electronic circuits (**Fig. 3a**). Here, our SAW device is driving the gate of a GaN HEMT. While the GaN HEMT features an ultrafast switching speed and low conduction loss, a sufficiently fast gate driver is necessary to fully exploit its performance.

To match the recommended gate-source voltage of the GaN HEMT, we use an input RF power of 34 dBm (2.5 W), resulting in an output voltage of 6.23 V. The 223-MHz RF signal is modulated by two square-wave pulses with 10-μs durations and a 10-μs separation, turning the RF signal fully on and off. This yields a rectified square-wave gate-source voltage $V_{GS}$ waveform (**Fig. 3b**). When the RF input is on, gate capacitor $C_G$ is being charged exceeding the threshold gate voltage, and the GaN HEMT is turned on, leading to a near-zero drain-source voltage $V_{DS}$. When the RF input is off, the gate resistor $R_G$ discharges the gate capacitor, and the GaN HEMT is turned off, leading to a large $V_{DS}$ (**Fig. 3c**). We note that although here we use our SAW gate driver in a unipolar configuration, push-pull structure can also be realized for applications requiring bipolar voltages or active control of turn-off.

During the first $V_{GS}$ pulse, the GaN HEMT is fully on, and drain-source current $I_{DS}$ increases linearly (**Fig. 3d**). The current increasing rate $\frac{dI_{DC}}{dt} = \frac{V_{DC}-V_{DS}}{L} \approx \frac{V_{DC}}{L}$, where $V_{DC}$ is the voltage of the DC power supply, $L$ is the inductance of the inductor, and $V_{DS} \approx 0$ when GaN HEMT is turned on. At the end of the first pulse, the HEMT turns off with a rapid drop in $I_{DS}$ and rise in $V_{DS}$. We extract the hard (with large $I_{DS}$) turn-off time $t_{OFF}$ of 121.3 ns (**Fig. 3e**). This turn-off time is primarily governed by the discharge of the gate capacitance through the gate resistor $R_G$, Between the interval of two pulses, the inductor maintains the current by circulating it through a silicon carbide Schottky freewheeling diode (FWD). At the start of the second $V_{GS}$ pulse, the GaN HEMT turns on at the current level of 1.65 A. The overshoot current at the beginning of the pulse (near time of 25.6 µs in **Fig. 3d**) is caused by the small reverse current due to the junction-capacitance-related transient of the FWD.

We measured a hard turn-on time $t_{ON}$ of 108.8 ns (**Fig. 3f**). Moreover, the SAW device exhibits a linear increase in output voltage regarding input RF power (**Fig. 3g**), which also improves the GaN HEMT turn-on speed at higher RF power levels (**Fig. 3h**). We note that the turn-on time is determined by the charging current and bandwidth of SAW gate driver. When the RF power is low (for example, <2 W), the charging current limits the turn-on time. On the other hand, when RF power is sufficiently large, the turn-on time will be more limited by the bandwidth of the IDTs. Achieving a 100-ns turn-on time demonstrates that our SAW gate driver can support 100-kHz switching applications. Further optimization of our SAW devices to achieve higher power delivery with larger bandwidth is needed for higher MHz-switching applications.

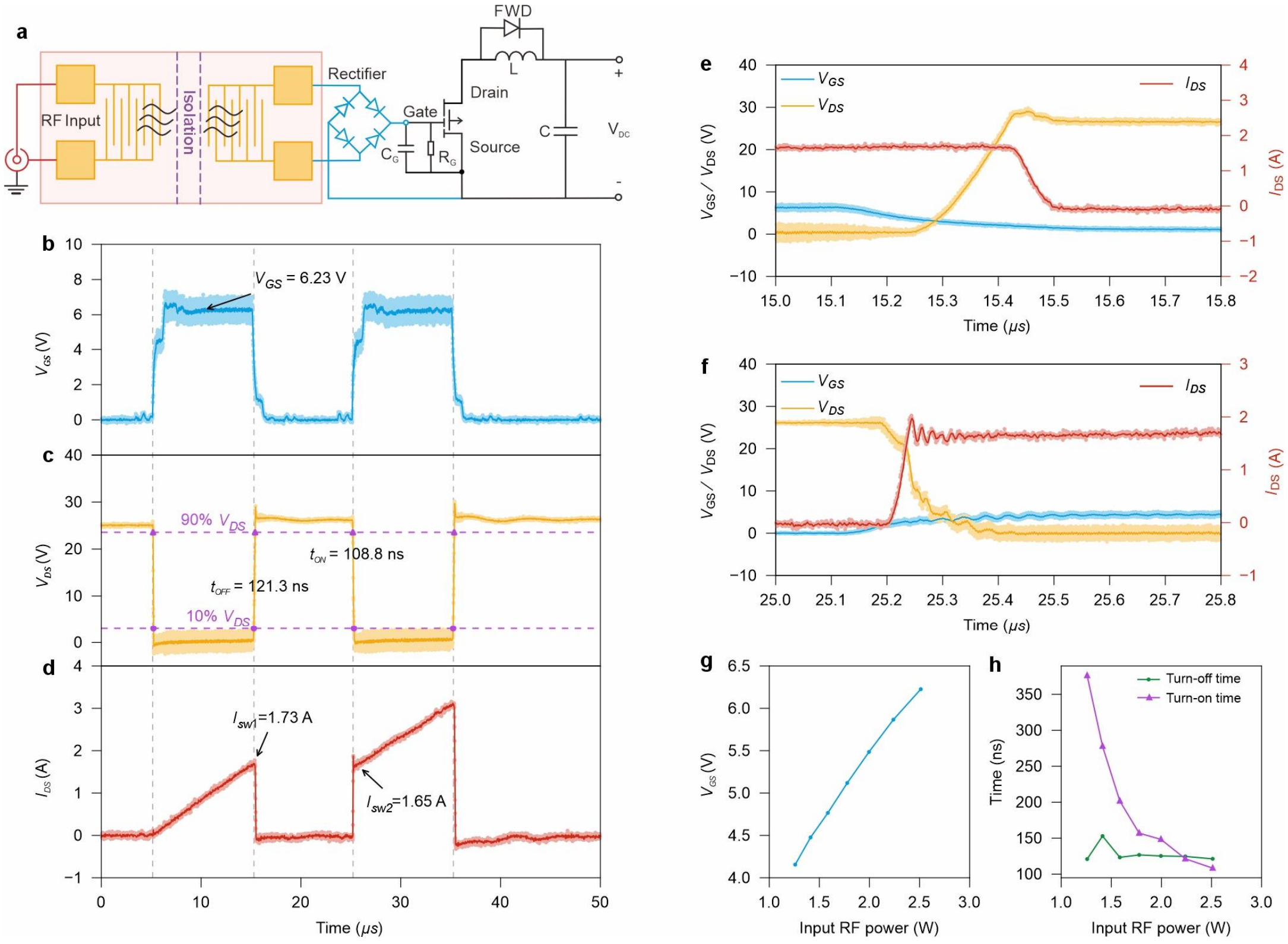

**Fig. 3 | Microwave-acoustic-driven GaN HEMT switching dynamics. a,** Circuit schematic of DPT for a GaN HEMT, $C_G$ discharged by a resistor during turn-off transient. GaN HEMT switching waveforms of **b,** gate-source voltage ($V_{GS}$). **c,** drain-source voltage ($V_{DS}$). **d,** current ($I_{DS}$). **e,** GaN HEMT turnoff dynamics. **f,** GaN HEMT turn-on dynamics. **g,** Gate voltage versus input RF power. **h,** Swiching times versus input RF power.

The ultimate metric for a gate driver's performance is its practical impact on the transistor's switching efficiency. Therefore, we further quantified the switching energy losses of the GaN HEMT our SAW driver. We calculate the switching loss by performing integral of drain-to-source voltage and current. Energy losses during hard turn-off and turn-on transient are 2.48 μJ and 1.43 μJ per switch, respectively. Most energy dissipates in the Miller plateau[37] duration of the switching process, when the gate driver current is diverted to charge the gate-to-drain capacitor. Therefore, switching losses can be significantly reduced if the gate driver circuit can provide more current to shorten the Miller capacitor charging time. With its tunable output voltage and current, the SAW gate driver could apply optimized gate voltages during different turn-on stages to minimize switching loss while suppressing switching ringing in the future. The turn-off losses depend on the GaN HEMT turnoff time, *i.e.*, how fast the gate capacitor can discharge through the resistor rather than the gate driver's performance. In other words, the input RF power has minimal impact on the GaN HEMT turn-off time.

To further reduce the turn-off time, we add a forward diode and a PNP transistor to our gate driving circuit (**Fig. 4a**). During the turn-off, the low voltage at our SAW gate driver's output pulls the base of the PNP transistor, which forward-biases its emitter-base junction. This results in an amplified discharge current from GaN HEMT's gate to source, enabling the fastest possible turn-off process. Remarkably, the fast turn-off and relatively slower turn-on are desired in power electronic applications, as fast turn-off reduces turn-off loss while a relatively slow turn-on suppresses common-mode noise and mitigates voltage overshoot or ringing.

### SAW driven DC-DC converter operation

Using our enhanced SAW gate driver circuit, we demonstrate a buck converter (**Fig. 4a**), representing the most common topology in power electronics systems. We create a buck converter using GaN HEMTs, achieving a 15 V ($V_{in}$) to 5.53 V ($V_{out}$) voltage step-down conversion. The high-side HEMT is driven by a 50 kHz PWM signal (50% duty cycle) from our isolated SAW gate driver. When the gate voltage is at logic high, the GaN HEMT turns on. The steady-state inductor voltage is approximately $v_L = V_{in} - V_{out} = L\frac{di_L}{dt}$, yielding a linear increase in the inductor current. When the gate voltage is at logic low, the GaN HEMT turns off. The magnetic energy stored in the inductor keeps the current circulating via the bottom diode. If the diode threshold voltage is negligible, the inductor voltage during the turnoff stage is $v_L = -V_{out} = Ldi_L/dt$ , causing the inductor current to decrease with a negative slope. The textbook like buck converter waveforms shows the high-performance high-side floating driving by our SAW gate driver, achieving a 5.53 V DC output with an efficiency of 73.7%. The conversion ratio $V_{out}/V_{in}$ and efficiency are limited by the excessive conduction loss during the current freewheeling stage through the low-side diode, which is implemented by another HEMT with 0 V gate-to-source voltage in this demonstration. The breakdown of losses is shown in **Supplementary Table 1**. This loss could be reduced by active driving both the high- and low-side HEMTs.

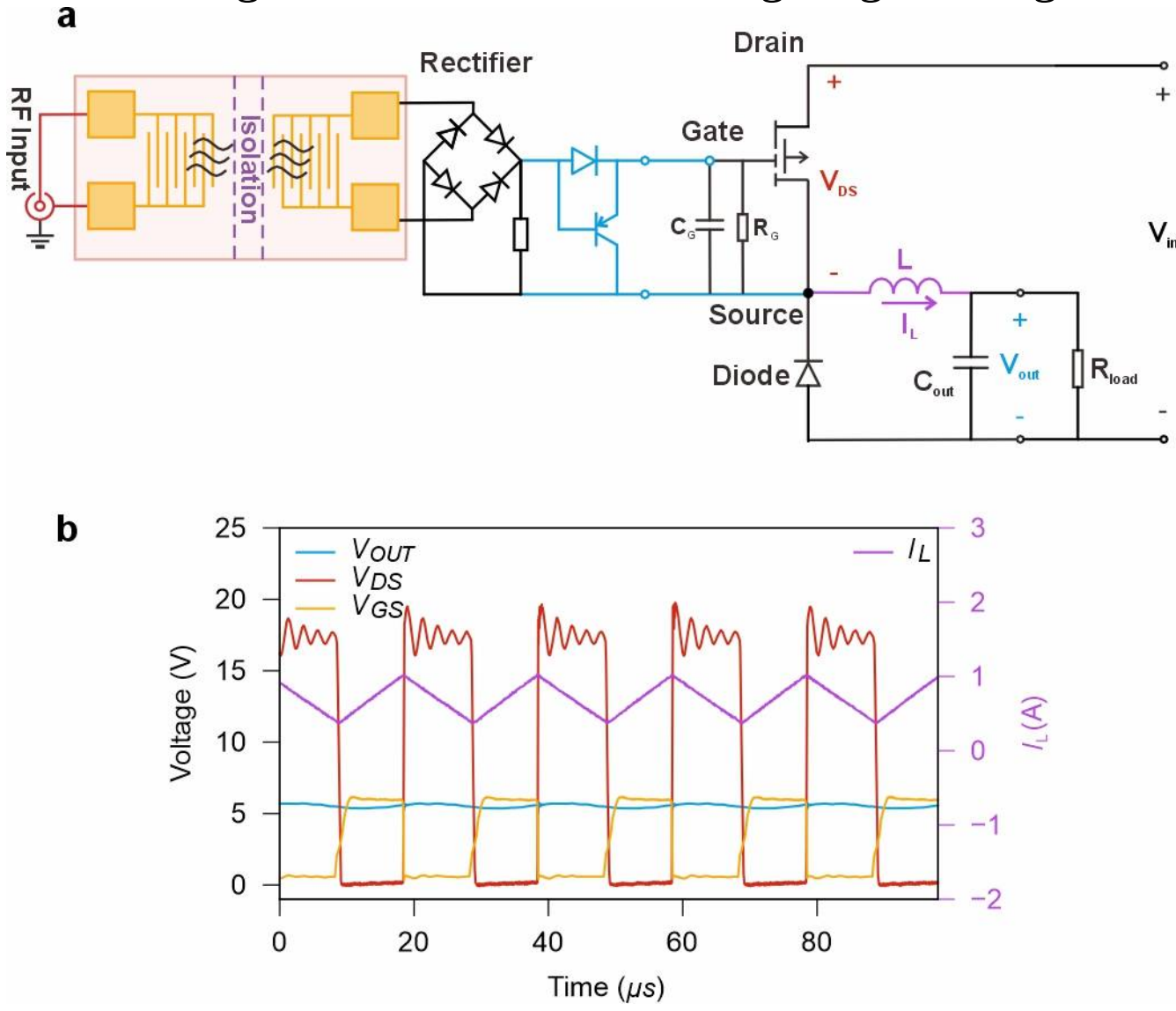

**Fig. 4 | SAW driven DC-DC converter with enhanced gate driver circuit. a,** Circuit schematic of our SAW-driven buck converter and **b,** Operation waveforms of output voltage $V_{out}$, drain-source voltage $V_{DS}$ and gate-source voltage $V_{GS}$.

### Wide temperature range operation of SAW device

Our SAW device features a wide operational temperature range (**Fig. 5**) from cryogenic temperature (0.535 K, -272.6 °C) to high temperature (544 K, 271 °C), which is much wider than deep space-qualified electronics (from –55 °C to 125 °C) as well as extreme temperatures on the moon (25 K to 410 K)[33]. Cryogenic power systems are needed for large-scale quantum computing systems[38-40], quantum-limited detectors[41,42], and space applications[43-45]. Meanwhile, high temperature operation is desired for many industrial applications, such as, sensors and control circuits near engines. The output gate driving voltage increases from 4.5 V to 5.6 V as temperature decreases from room temperature 294.7 K to 0.535 K (**Fig. 5a**), as the propagation loss of SAW is much smaller at cryogenic temperatures than that at room temperature.

In high temperature measurements, the output voltage remains stable when the SAW device operates in air, from 290 K (17 °C) to 473 K (200 °C) (**Fig. 5b**). Our SAW device shows better overall performance at cryogenic environment as the propagation loss of SAW is much smaller at cryogenic temperatures than that at higher temperatures (room temperature to 200 °C)[33]. At temperatures above 473 K (200 °C), we observe the output voltage begins to degrade; we speculate the degradation is due to the melting of the solder used on the mounting printed circuit board. We note the piezoelectricity of LN was tested stable from room temperature to 750 °C[46], and the temperature coefficient of acoustic propagation loss is estimated at $4 \times 10^{-4}$ dB cm$^{-1}$ °C$^{-1}$ for 200 MHz SAW on LN.

Notably, the rise and fall times of the output gate driving voltage remain the same over the temperature range. Focusing on our SAW device in this work, we only test the SAW device at different temperatures, and the diode rectifier remains at room temperature. Although silicon diode we used in this work are not rated for such wide temperature range, GaN HEMT devices have been demonstrated operational down to 10 mK[47]. Meanwhile, SiC, gallium oxide ($Ga_2O_3$), and diamond devices have been demonstrated up to 573 K (300 °C)[48], 600 K (327 °C)[49], and 1273 K (1000 °C)[50], respectively. Future circuit integration with these devices could provide a pathway toward compact SAW gate drivers suitable for ultrawide operational temperature range applications.

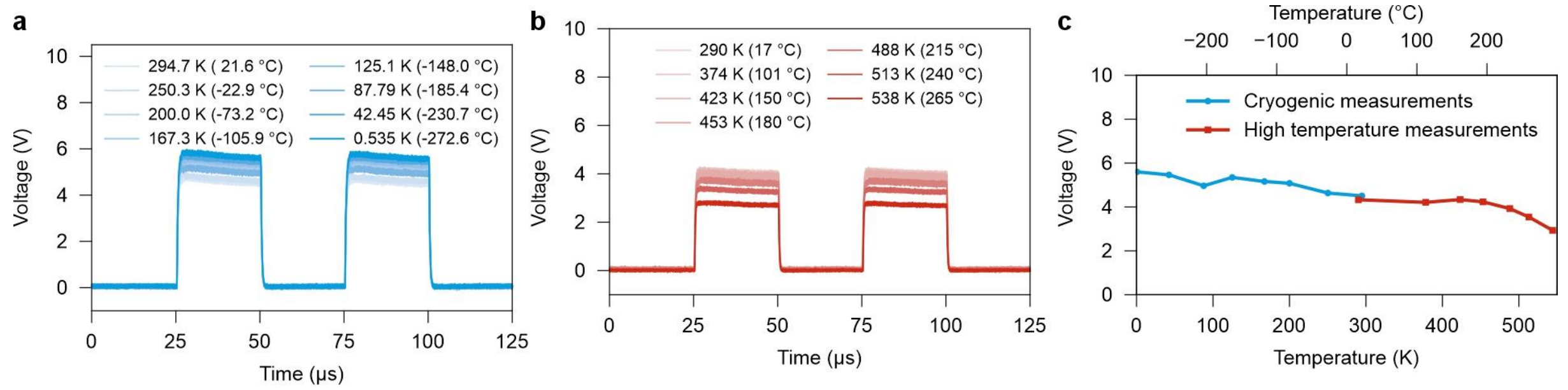

**Fig. 5 | Wide temperature range operation of our SAW device. a,** Cryogenic temperature operation waveforms down to 0.535 K. **b,** High temperature operation waveforms up to 544 K (271 °C). **c,** SAW device output peak voltage versus temperature.

### Discussion

The choice of acoustic-wave frequency around 200 MHz balances the instantaneous bandwidth and power handling capabilities, which are two main factors limiting the gate driving performance. A higher frequency device can lead to larger instantaneous bandwidth, meanwhile its smaller pitch of electrode limits the maximum voltage can be applied at input. On LN substrate, a typical device zero-to-zero bandwidth is about 10% (corresponding to a 3-dB bandwidth of 5%) of the carrier frequency for efficient IDTs (as a result of the electromechanical coupling coefficient $k^2 \sim 5.3\%$). As a ferroelectric material, the coercive electrical field of LN is about 20 V/μm. Our 223 MHz SAW device (with the electrode pitch of 8.55 μm) lead to a sub-100 ns rise/fall time, suitable for gate driving of GaN power electronic HEMT. Although the current SAW gate driver performance has not reached GaN devices' full potential of MHz ultrafast switching, it is promising to reduce the switching times by adopting higher-frequency, broader-bandwidth SAW designs with reduced insertion loss in the future. In addition, our devices could be optimized for different power transistors: For example, compared to the GaN HEMT used in this work, a SAW gate driver with a higher gate driving voltage is preferred for high-voltage high-current SiC MOSFET. **Supplementary Table 2** compares our SAW gate driver with other isolated gate drivers.

## Conclusions and outlook

In summary, we establish isolated power and signal co-transmission approach based on integrated microwave-frequency SAW devices. Our microwave-acoustic device directly confronts the limitations of existing technologies. First, using mechanical waves as the transmission medium could avoid electrical interference from both radiative and conductive EMI. Second, it unifies the power and signal paths, eliminating the need for a separate isolated power supply and its associated components. Third, microwave-SAW devices support the modulation of multi-megahertz PWM signals, meeting the demands of next-generation power electronics and boarder applications. Acoustic-wave devices could be fully integrable with multiple semiconductor platforms, including silicon[51-53], silicon carbide[54,55], GaN[56-58], and diamond[59]. Specifically, transferred thin-film LN on diamond[59] has demonstrated without compromising acoustic-wave performance; chip die with piezoelectric devices has been 3D integrated with RF circuits[53]. We believe it is feasible to integrated acoustic-wave devices with power devices, providing robust on-chip electrical isolation for truly monolithic, intelligent power semiconductors.

## Methods

### SAW Device fabrication and characterization

Our device is fabricated on 128°Y cut black LN substrate, and the SAW is propagating in the crystal X direction. The IDT aperture is 500 μm, with 20 pairs of finger electrodes distributed in a pitch of 8.55 μm, corresponding to an acoustic frequency of 223 MHz. The IDT contact pads are 300 μm × 300 μm. The device is patterned using a maskless aligner (Heidelberg Instruments MLA 150) using photoresist (Microposit S1818), followed by electron beam evaporation of 10-nm-thick chromium and 90-nm-thick gold, and lift-off process using Microposit Remover 1165. The cross-section schematic of our IDT design is shown in Supplementary Fig.1.

The 223-MHz input RF signals are generated by a vector signal generator (Rohde & Schwarz SMW200A). The RF signals are then modulated by square waves from an arbitrary waveform generator (Rigol DG 852 Pro). The modulated signals are amplified by 53 dB using a high power RF amplifier (Mini-circuits ZHL-10M2G0020+) before delivering to the input port of our SAW device.

The microwave performance of the SAW devices at room temperature is characterized using a home-built probe station with microwave instruments. We use a network analyzer (Keysight 5000A or LibreVNA 2.0) to measure the $S$ parameter (**Fig. 2b**) and to extract isolation capacitance (**Fig. 2h**). Measurements are calibrated using a standard calibration substrate (GGB CS-108). The isolation capacitance is extracted from the calibrated $S$ parameter, averaging from 80 to 85 MHz. A source measure unit (Keysight B2912A) is used to measure the output characterization (**Fig. 2e**). The breakdown voltage (**Fig. 2g**) is measured by a Keysight B1505 curve tracer.

### Gate driving circuit

A full-bridge rectifier is used to down convert the modulated RF signals to low-frequency gate driving signals. The full-bridge rectifier includes two dual Schottky diode arrays (Infineon BAS4004E6237HTSA1), a capacitor, and a load resistor are added to hold the voltage at output and allows discharge of gate voltage. A 72-pF capacitor and 1-kΩ resistor are used in device characterization (**Fig. 2**). To match the optimal rated gate voltage of 6 to 7 V of the GaN HEMT, the resistor value is adjusted to 385 Ω (**Fig. 3**).

In the enhanced gate driving circuit design (**Fig. 4**), a local PNP transistor and a forward diode are added in cascade between the rectifier and the HEMT gate for accelerating the gate discharging process. During the turnoff process, the PNP transistor shorts the gate to source to create a stronger discharging path while the forward diode has limited impact on the turn-on speed.

## Double pulse test

Our DPT setup (**Fig. 2a**) includes an input DC power supply (Rigol DP831A) of 25 V, a DC-link capacitor (C) of 94 μF, an inductor (L) of 132 μH and a SiC Schottky freewheeling diode (FWD, C4D10120). The device under test (DUT) is a 650-V, 11-A GaN power HEMT, model GS-065-011-1L. Utilizing the double pulse test, we evaluate the turn-on/turnoff time and the switching energy losses of the DUT with negligible thermal effect. The turn-on and turnoff times are measured when $V_{DS}$ transits between 10% and 90% of its peak value.

## DC-DC buck converter

We build our buck converter using two GS-065-011-1L GaN HEMTs (**Fig. 4a**). Also, a low-pass filter, including a 132 μH inductor (L), 4.2 μF capacitor ($C_{out}$), enables DC output to an 8 Ω electronic load ($R_{load}$, Rigol DL3031). The DC power supply of 15 V is provided by Rigol DP831A. The high-side HEMT is gate controlled by our SAW device, switching on and off at 50 kHz with 50% duty cycle. We short the gate and source of the low-side HEMT. When the high side HEMT is off, the inductor current can still be reversely conducted by the low-side HEMT working in the third quadrant. However, the low side HEMT presents higher drain-to-source voltage drop in reverse conduction mode during converter operation. This explains the relatively low efficiency of 73.7% of our converter.

## Cryogenic and high temperature measurements

We wire bond our SAW device on a printed circuit board (PCB) for cryogenic and high temperature measurements. For device performance characterization at cryogenic temperatures, we mount the PCB on the still plate of a dilution fridge (Bluefors) with a base temperature below 1 K. The amplified RF signal is applied to the SAW device through the superconducting coaxial cables of the dilution fridge. We cool down the fridge from room temperature (295 K) to 0.535 K and run the measurements for 24 hours. The temperature is read out by a temperature sensor on the mounting plate in the dilution fridge. For high temperature experiment, we mount the PCB on a hot plate to increase the temperature from 17 °C to 271 °C. The temperature is monitored by a commercial thermal camera with a resolution of 0.1 °C. At 271 °C, we observe the solder used on the PCB starts melting. **Supplementary Fig. 4** shows experimental setups and configurations in these measurements.

## Acknowledgements

We thank Prof. Thomas O'Donnell for the wire bonder and Rohde & Schwarz for support with microwave instrumentation. This project was partially supported by faculty start-up funding (L.S., L.Z.) from Bradley Department of Electrical and Computer Engineering of Virginia Tech. Device fabrication was conducted at the Center for Nanophase Materials Sciences (CNMS2024-B-02643, L.S.), which is a US Department of Energy Office of Science User Facility. Development of the optical vibrometer was partially supported by the Defense Advanced Research Projects Agency (DARPA) OPTIM program (HR00112320031, L.S. and Y.Z.). The views and conclusions contained in this document are those of the authors and do not necessarily reflect the position or the policy of the United States Government. No official endorsement should be inferred. Approved for public release.

## Author contributions statement

L.S. and L.Z. conceptualized the idea. L.S., Z.X., and J.J. fabricated the devices. L.J., L.S., and L.Z. performed the room-temperature SAW device characterization and gate driving experiments. L.J., Y. Zhang, and N.C. performed the cryogenic measurements. J.G.T. and Y. Zhu performed the optical vibrometry of the devices. L.J., L.S. and L.Z. analyzed the data. L.J. prepared the draft of the manuscript. L.S. and L.Z. revised the manuscript. L.S. and L.Z. supervised the project.

## Competing interests

Virginia Tech has filed a patent application based on this work. The authors declare no other competing interests.

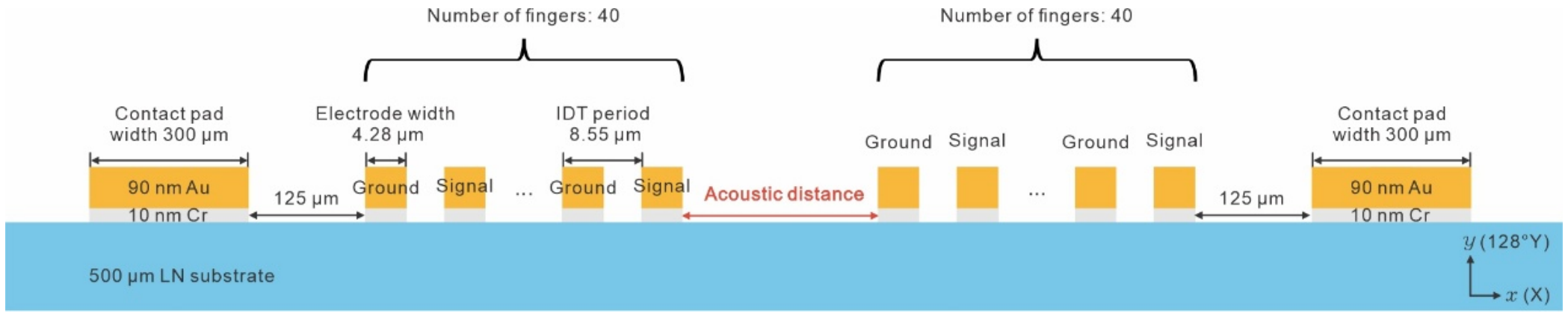

**Supplementary Fig. 1 |** Cross-section schematic of our IDT design. Our device is fabricated on 128°Y cut black lithium niobate (LN) substrate with 500 μm thickness, and the SAW is propagating in the crystal X direction. The IDT contact pads are 300 μm × 300 μm. The IDT aperture at both input and output port is 500 μm, with 20 pairs of finger electrodes distributed in a pitch of 8.55 μm, corresponding to an acoustic frequency of 223 MHz. The electrodes consist of 10-nm-thick chromium and 90-nm-thick gold.

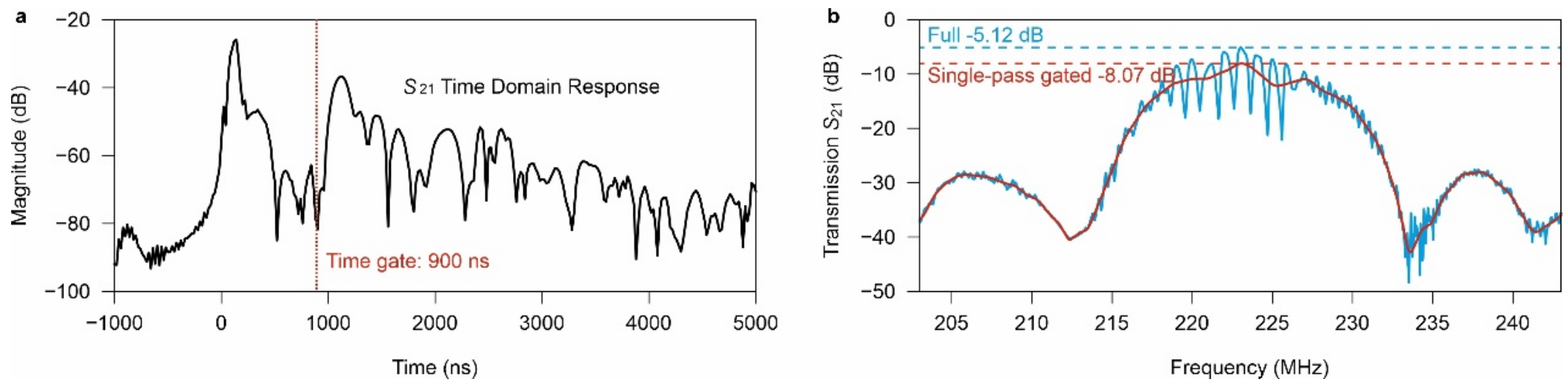

**Supplementary Fig. 2 | a,** Time domain response of $S_{21}$ transmission spectrum**. b,** $S_{21}$ spectra of total IDT transmission and gated single-pass acoustic-wave transmission.

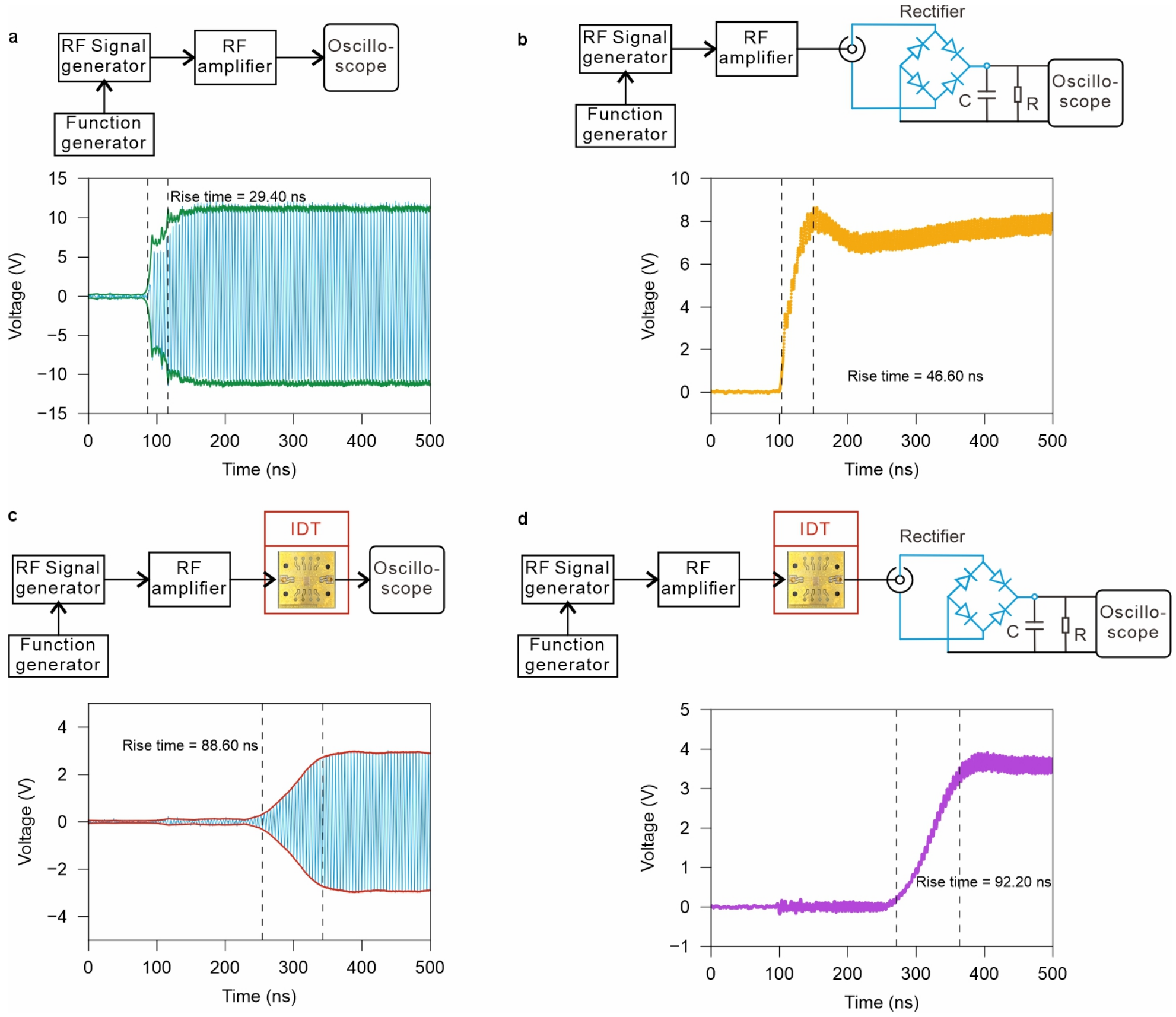

**Supplementary Fig. 3 | Rise time of signal in different test conditions. a,** Amplified modulated RF signal. **b,** after the rectifier. **c,** after the SAW device. **d,** after the SAW device and the rectifier. These tests use a different device than that shown in Fig. 2.

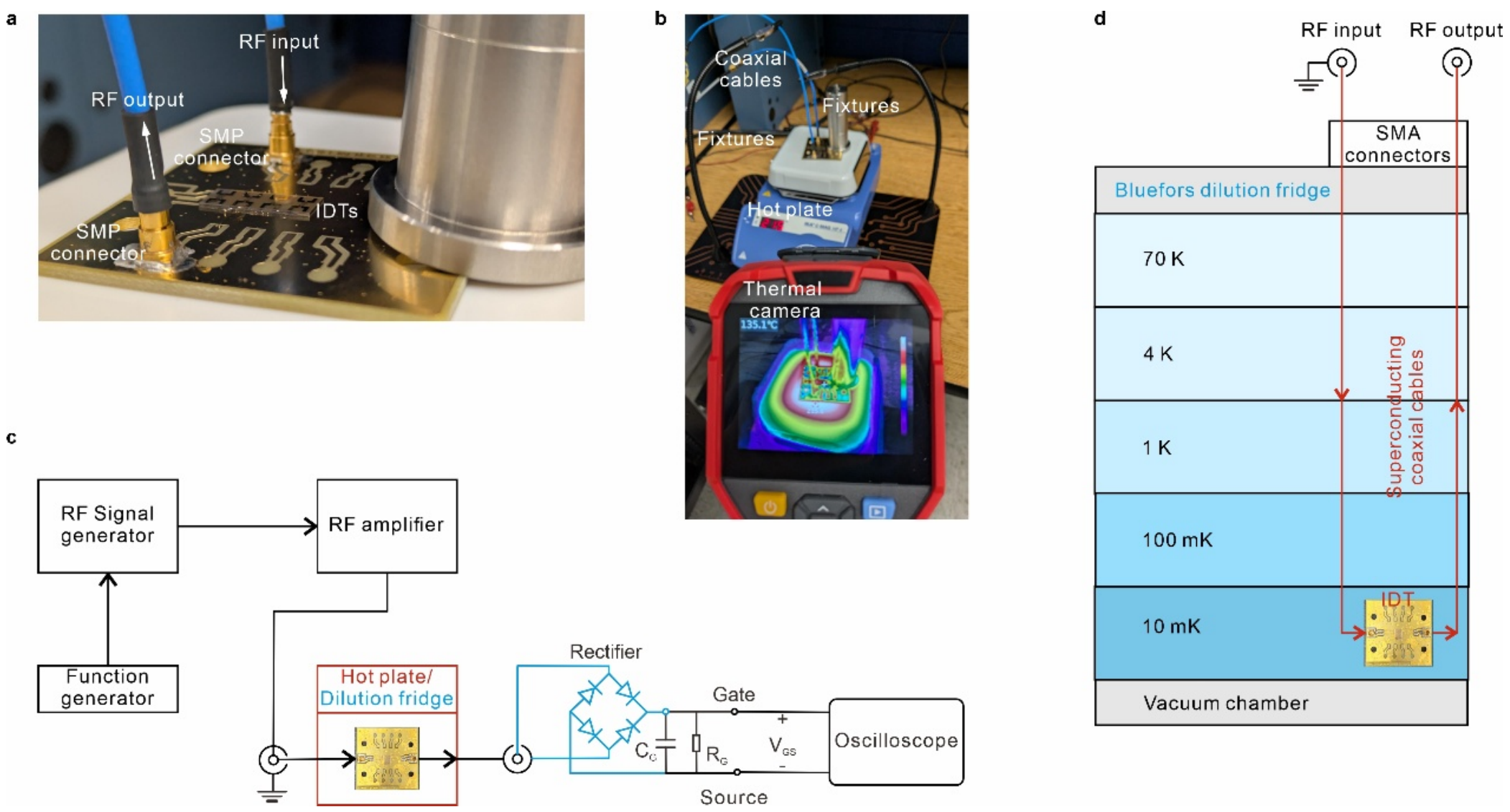

**Supplementary Fig. 4 | a,** Packaged IDT device on a PCB with SMP connectors. **b,** High-temperature test photographs. **c,** High-temperature gate driver test setup diagram. **d,** Cable connection in the dilution fridge.

**Supplementary Table 1. Breakdown of loss in the buck converter experiment.**

| Loss term | Symbol | Value (W) |
|---|---|---|
| Low side HEMT third-quadrant conduction loss | $P_{low_3rd}$ | 0.915 |
| High side HEMT switching loss | $P_{high_sw}$ | 0.239 |
| High side conduction loss | $P_{high_cond}$ | 0.034 |
| Inductor conduction loss | $P_L$ | 0.179 |
| $C_{oss}$ | $P_{oss}$ | 0.005 |
| Total loss | $P_{total}$ | 1.372 |

**Supplementary Table 2. | Comparison of isolated gate drivers for power electronics.**

| Reference | Isolated gate driver mechanism | Isolation capacitance (pF) | Breakdown voltage (V) | Operational bandwidth |
|---|---|---|---|---|
| Our SAW gate driver | Electromechanical conversion | 0.032 | 2750 | 4.9 MHz |
| [1] | Inductive coupling | 2.20 | 25000 | 4 MHz |
| Infineon 1EDC20I12MH | | - | 1000 | 1 MHz |
| [2] | Capacitive coupling | 0.3 and 0.7 | - | < 200 kHz |
| Texas Instrument UCC21225A | | 1.2 | 2500 | 5 MHz |
| [3] | Optical gate driver | - | - | <10 MHz |